\documentclass{article}
\usepackage{spconf,amsmath,graphicx,url,bm,array,xcolor,booktabs, blindtext, enumitem, multirow}
\setlist[description]{leftmargin=0cm,labelindent=0cm}
\usepackage[ruled,vlined, linesnumbered]{algorithm2e}
\usepackage[
	n,
	operators,
	advantage,
	sets,
	adversary,
	landau,
	probability,
	notions,	
	logic,
	ff,
	mm,
	primitives,
	events,
	complexity,
	asymptotics,
	keys]{cryptocode}
\usepackage{csquotes}
\usepackage{dashbox}
\usepackage{todonotes}
\usepackage{listings}
\usepackage{trace}
\usepackage{makeidx}
\usepackage{mdframed}

\let\oldnl\nl% Store \nl in \oldnl
\newcommand{\nonl}{\renewcommand{\nl}{\let\nl\oldnl}}% Remove line number for one line
\def\x{{\mathbf x}}

\title{Encrypted Speech Recognition using Deep Polynomial Networks}
\name{Shi-Xiong Zhang,  Yifan Gong$^\dagger$ and Dong Yu}
\address{Tencent AI Lab, Bellevue, USA \\
$^\dagger$Microsoft Corporation, Bellevue, USA \\
{\small \tt \{auszhang,dyu\}@tencent.com, ygong@microsoft.com}}
\begin{document}
%\ninept
%
\maketitle
\begin{abstract}
The cloud-based speech recognition/API provides developers or enterprises an easy way to create speech-enabled features in their applications. However, sending audios about personal or company internal information to the cloud, raises concerns about the privacy and security issues. The recognition results generated in cloud may also reveal some sensitive information. This paper proposes a deep polynomial network (DPN) that can be applied to the encrypted speech as an acoustic model. It allows clients to send their data in an encrypted form to the cloud to ensure that their data remains confidential, at mean while the DPN can still make frame-level predictions over the encrypted speech and return them in encrypted form. One good property of the DPN is that it can be trained on unencrypted speech features in the traditional way. To keep the cloud away from the raw audio and recognition results, a cloud-local joint decoding framework is also proposed. We demonstrate the effectiveness of model and framework on the Switchboard and Cortana voice assistant tasks with small performance degradation and latency increased comparing with the traditional cloud-based DNNs. 
\end{abstract}
\begin{keywords}
speech recognition, privacy preserving, encryption, DNN, quantization 
\end{keywords}
\section{Introduction}
\label{sec:intro}

Modern speech recognition services are running on cloud \cite{huang2014historical}. These cloud-based speech-to-text API provides developers or enterprises an easy way to create powerful speech-enabled features in their own applications like voice-based command control, user dialog using natural speech conversation, and speech transcription and dictation. However, in the scenarios that involve medical, financial, or enterprise sensitive data, applying cloud-based speech recognition might be prohibited due to the privacy and legal requirements regarding the confidentiality of the information \cite{pathak2013privacy,smaragdis2007framework, hesamifard2018privacy, miotto2017deep, CryptoNets}. 

In this work we present a practical framework to allow third parties to use speech recognition services without sacrificing their privacy. In the proposed framework, the audio provider first encrypts the features extracted from the audio in local, then sends the features in encrypted form to the acoustic model on the cloud. We proposed a special form of DNNs, deep polynomial networks (DPN), as acoustic models, which can make predictions over the encrypted input features and yields the posteriors also in the encrypted form. % that only the original audio provider can decrypt and decode
Note that the server does not have the private key, so it is impossible for the server to decrypt the frame-level scores (posteriors) \cite{ CryptoNets, dowlin2017manual}.
%nor decode the speech\footnote{Viterbi algorithm} \cite{dowlin2017manual, CryptoNets}. 
Also to keep the cloud server away from the final recognition results, a cloud-local joint decoding framework is proposed. 
In this framework the encrypted frame-level scores from cloud will be sent back to the local owner who has the secret key for decryption. The decoding is done in local after the decryption. During the entire process no information is exposed to the cloud. It just performed a frame-level prediction on behalf of audio providers and learn nothing about the customer's data or the recognition results.

The main contributions of this work include: $1$) a cloud-local joint decoding framework that enables practical privacy preserving speech recognition.
%that keeps the accuracy of DNNs with the simplicity of use of homomorphic encryption. 
$2$) a deep polynomial network that can be trained on unencrypted data and make predictions over the encrypted inputs in real time. The remainder of the paper is organized as follows. In Section \ref{sec:Frame} we introduce the homomorphic encryption and the challenges of applying it to speech recognition. In Section \ref{sec:PolyNet} we describe the structures of polynomial networks for efficient homomorphic encryption. The experimental results will be discussed in Section \ref{sec:Exp}.

\section{Encrypted Speech Recognition}
\label{sec:Frame}

\subsection{Homomorphic Encryption}
\label{ssec:HE}

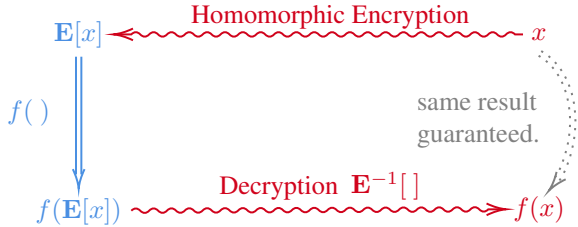
\begin{figure}[t]
  %\centerline{\includegraphics[width=8cm]{HE.png}} 

\tikzset{every picture/.style={line width=0.75pt}} %set default line width to 0.75pt        

\begin{tikzpicture}[x=0.75pt,y=0.75pt,yscale=-.9,xscale=.9]

%Curve Lines [id:da23447474465779838] 
\draw [color={rgb, 255:red, 128; green, 128; blue, 128 }  ,draw opacity=1 ] [dash pattern={on 0.84pt off 2.51pt}]  (409.2,115.36) .. controls (421.41,122.2) and (428.09,137.23) .. (428.35,153.74) .. controls (428.36,154.07) and (428.36,154.41) .. (428.36,154.74) .. controls (428.36,157.66) and (428.16,160.62) .. (427.76,163.58) .. controls (426.34,173.97) and (422.42,184.36) .. (419.13,188.22)(407.73,117.98) .. controls (419.13,124.36) and (425.11,138.49) .. (425.35,153.79) .. controls (425.36,154.1) and (425.36,154.42) .. (425.36,154.74) .. controls (425.36,157.53) and (425.17,160.35) .. (424.78,163.18) .. controls (423.44,173.07) and (419.71,182.97) .. (416.74,186.37) ;
\draw [shift={(413.47,193.67)}, rotate = 309.03999999999996] [color={rgb, 255:red, 128; green, 128; blue, 128 }  ,draw opacity=1 ][line width=0.75]    (10.93,-3.29) .. controls (6.95,-1.4) and (3.31,-0.3) .. (0,0) .. controls (3.31,0.3) and (6.95,1.4) .. (10.93,3.29)   ;

% Text Node
\draw (150,106) node   {$\mathbf{\textcolor[rgb]{0.29,0.56,0.89}{E}}\textcolor[rgb]{0.29,0.56,0.89}{[}\textcolor[rgb]{0.29,0.56,0.89}{\mathnormal{x}}\textcolor[rgb]{0.29,0.56,0.89}{]}$};
% Text Node
\draw (150,205) node   {$\mathnormal{\textcolor[rgb]{0.29,0.56,0.89}{f}\textcolor[rgb]{0.29,0.56,0.89}{(}}\mathbf{\textcolor[rgb]{0.29,0.56,0.89}{E}}\textcolor[rgb]{0.29,0.56,0.89}{[}\textcolor[rgb]{0.29,0.56,0.89}{\mathnormal{x}}\textcolor[rgb]{0.29,0.56,0.89}{]}\textcolor[rgb]{0.29,0.56,0.89}{)}$};
% Text Node
\draw (408,106) node   {$\textcolor[rgb]{0.82,0.01,0.11}{\mathnormal{x}}$};
% Text Node
\draw (408,204) node   {$\mathnormal{\textcolor[rgb]{0.82,0.01,0.11}{f}\textcolor[rgb]{0.82,0.01,0.11}{(}\textcolor[rgb]{0.82,0.01,0.11}{x}\textcolor[rgb]{0.82,0.01,0.11}{)}}$};
% Text Node
\draw (122,150) node   {$\mathnormal{\textcolor[rgb]{0.29,0.56,0.89}{f}\textcolor[rgb]{0.29,0.56,0.89}{(}\textcolor[rgb]{0.29,0.56,0.89}{\ }\textcolor[rgb]{0.29,0.56,0.89}{)}}$};
% Text Node
\draw (291,96) node  [align=left] {\textcolor[rgb]{0.82,0.01,0.11}{Homomorphic Encryption}};
% Text Node
\draw (375,170) node  [align=left] {\textcolor[rgb]{0.5,0.5,0.5}{same result }\\\textcolor[rgb]{0.5,0.5,0.5}{guaranteed.}\\\\};
% Text Node
\draw (264,192) node  [align=left] {\textcolor[rgb]{0.82,0.01,0.11}{Decryption }};
% Text Node
\draw (322,191) node   {$\mathbf{\textcolor[rgb]{0.82,0.01,0.11}{E}}\textcolor[rgb]{0.82,0.01,0.11}{^{-1}}\textcolor[rgb]{0.82,0.01,0.11}{[}\textcolor[rgb]{0.82,0.01,0.11}{\ }\textcolor[rgb]{0.82,0.01,0.11}{]}$};
% Connection
\draw [color={rgb, 255:red, 208; green, 2; blue, 27 }  ,draw opacity=1 ]   (171.5,106) -- (179.5,106) .. controls (181.17,104.33) and (182.83,104.33) .. (184.5,106) .. controls (186.17,107.67) and (187.83,107.67) .. (189.5,106) .. controls (191.17,104.33) and (192.83,104.33) .. (194.5,106) .. controls (196.17,107.67) and (197.83,107.67) .. (199.5,106) .. controls (201.17,104.33) and (202.83,104.33) .. (204.5,106) .. controls (206.17,107.67) and (207.83,107.67) .. (209.5,106) .. controls (211.17,104.33) and (212.83,104.33) .. (214.5,106) .. controls (216.17,107.67) and (217.83,107.67) .. (219.5,106) .. controls (221.17,104.33) and (222.83,104.33) .. (224.5,106) .. controls (226.17,107.67) and (227.83,107.67) .. (229.5,106) .. controls (231.17,104.33) and (232.83,104.33) .. (234.5,106) .. controls (236.17,107.67) and (237.83,107.67) .. (239.5,106) .. controls (241.17,104.33) and (242.83,104.33) .. (244.5,106) .. controls (246.17,107.67) and (247.83,107.67) .. (249.5,106) .. controls (251.17,104.33) and (252.83,104.33) .. (254.5,106) .. controls (256.17,107.67) and (257.83,107.67) .. (259.5,106) .. controls (261.17,104.33) and (262.83,104.33) .. (264.5,106) .. controls (266.17,107.67) and (267.83,107.67) .. (269.5,106) .. controls (271.17,104.33) and (272.83,104.33) .. (274.5,106) .. controls (276.17,107.67) and (277.83,107.67) .. (279.5,106) .. controls (281.17,104.33) and (282.83,104.33) .. (284.5,106) .. controls (286.17,107.67) and (287.83,107.67) .. (289.5,106) .. controls (291.17,104.33) and (292.83,104.33) .. (294.5,106) .. controls (296.17,107.67) and (297.83,107.67) .. (299.5,106) .. controls (301.17,104.33) and (302.83,104.33) .. (304.5,106) .. controls (306.17,107.67) and (307.83,107.67) .. (309.5,106) .. controls (311.17,104.33) and (312.83,104.33) .. (314.5,106) .. controls (316.17,107.67) and (317.83,107.67) .. (319.5,106) .. controls (321.17,104.33) and (322.83,104.33) .. (324.5,106) .. controls (326.17,107.67) and (327.83,107.67) .. (329.5,106) .. controls (331.17,104.33) and (332.83,104.33) .. (334.5,106) .. controls (336.17,107.67) and (337.83,107.67) .. (339.5,106) .. controls (341.17,104.33) and (342.83,104.33) .. (344.5,106) .. controls (346.17,107.67) and (347.83,107.67) .. (349.5,106) .. controls (351.17,104.33) and (352.83,104.33) .. (354.5,106) .. controls (356.17,107.67) and (357.83,107.67) .. (359.5,106) .. controls (361.17,104.33) and (362.83,104.33) .. (364.5,106) .. controls (366.17,107.67) and (367.83,107.67) .. (369.5,106) .. controls (371.17,104.33) and (372.83,104.33) .. (374.5,106) .. controls (376.17,107.67) and (377.83,107.67) .. (379.5,106) .. controls (381.17,104.33) and (382.83,104.33) .. (384.5,106) .. controls (386.17,107.67) and (387.83,107.67) .. (389.5,106) .. controls (391.17,104.33) and (392.83,104.33) .. (394.5,106) -- (399,106) -- (399,106) ;

\draw [shift={(169.5,106)}, rotate = 0] [color={rgb, 255:red, 208; green, 2; blue, 27 }  ,draw opacity=1 ][line width=0.75]    (10.93,-3.29) .. controls (6.95,-1.4) and (3.31,-0.3) .. (0,0) .. controls (3.31,0.3) and (6.95,1.4) .. (10.93,3.29)   ;
% Connection
\draw [color={rgb, 255:red, 74; green, 144; blue, 226 }  ,draw opacity=1 ]   (151.5,118.5) -- (151.5,184.5)(148.5,118.5) -- (148.5,184.5) ;
\draw [shift={(150,192.5)}, rotate = 270] [color={rgb, 255:red, 74; green, 144; blue, 226 }  ,draw opacity=1 ][line width=0.75]    (10.93,-3.29) .. controls (6.95,-1.4) and (3.31,-0.3) .. (0,0) .. controls (3.31,0.3) and (6.95,1.4) .. (10.93,3.29)   ;

% Connection
\draw [color={rgb, 255:red, 208; green, 2; blue, 27 }  ,draw opacity=1 ]   (179.5,204.89) .. controls (181.16,203.22) and (182.83,203.21) .. (184.5,204.87) .. controls (186.17,206.53) and (187.84,206.52) .. (189.5,204.85) .. controls (191.16,203.18) and (192.83,203.17) .. (194.5,204.83) .. controls (196.17,206.49) and (197.84,206.48) .. (199.5,204.81) .. controls (201.16,203.14) and (202.83,203.13) .. (204.5,204.79) .. controls (206.17,206.45) and (207.84,206.44) .. (209.5,204.77) .. controls (211.16,203.1) and (212.83,203.09) .. (214.5,204.75) .. controls (216.17,206.41) and (217.84,206.4) .. (219.5,204.73) .. controls (221.16,203.06) and (222.83,203.05) .. (224.5,204.71) .. controls (226.17,206.37) and (227.84,206.36) .. (229.5,204.69) .. controls (231.16,203.02) and (232.83,203.01) .. (234.5,204.67) .. controls (236.17,206.33) and (237.84,206.32) .. (239.5,204.65) .. controls (241.16,202.98) and (242.83,202.97) .. (244.5,204.63) .. controls (246.17,206.29) and (247.84,206.28) .. (249.5,204.61) .. controls (251.16,202.94) and (252.83,202.93) .. (254.5,204.59) .. controls (256.17,206.25) and (257.83,206.25) .. (259.5,204.58) .. controls (261.16,202.91) and (262.83,202.9) .. (264.5,204.56) .. controls (266.17,206.22) and (267.84,206.21) .. (269.5,204.54) .. controls (271.16,202.87) and (272.83,202.86) .. (274.5,204.52) .. controls (276.17,206.18) and (277.84,206.17) .. (279.5,204.5) .. controls (281.16,202.83) and (282.83,202.82) .. (284.5,204.48) .. controls (286.17,206.14) and (287.84,206.13) .. (289.5,204.46) .. controls (291.16,202.79) and (292.83,202.78) .. (294.5,204.44) .. controls (296.17,206.1) and (297.84,206.09) .. (299.5,204.42) .. controls (301.16,202.75) and (302.83,202.74) .. (304.5,204.4) .. controls (306.17,206.06) and (307.84,206.05) .. (309.5,204.38) .. controls (311.16,202.71) and (312.83,202.7) .. (314.5,204.36) .. controls (316.17,206.02) and (317.84,206.01) .. (319.5,204.34) .. controls (321.16,202.67) and (322.83,202.66) .. (324.5,204.32) .. controls (326.17,205.98) and (327.84,205.97) .. (329.5,204.3) .. controls (331.16,202.63) and (332.83,202.62) .. (334.5,204.28) .. controls (336.17,205.94) and (337.83,205.94) .. (339.5,204.27) .. controls (341.16,202.6) and (342.83,202.59) .. (344.5,204.25) .. controls (346.17,205.91) and (347.84,205.9) .. (349.5,204.23) .. controls (351.16,202.56) and (352.83,202.55) .. (354.5,204.21) .. controls (356.17,205.87) and (357.84,205.86) .. (359.5,204.19) .. controls (361.16,202.52) and (362.83,202.51) .. (364.5,204.17) .. controls (366.17,205.83) and (367.84,205.82) .. (369.5,204.15) .. controls (371.16,202.48) and (372.83,202.47) .. (374.5,204.13) .. controls (376.17,205.79) and (377.84,205.78) .. (379.5,204.11) -- (379.5,204.11) -- (387.5,204.08) ;
\draw [shift={(389.5,204.07)}, rotate = 539.78] [color={rgb, 255:red, 208; green, 2; blue, 27 }  ,draw opacity=1 ][line width=0.75]    (10.93,-3.29) .. controls (6.95,-1.4) and (3.31,-0.3) .. (0,0) .. controls (3.31,0.3) and (6.95,1.4) .. (10.93,3.29)   ;

\end{tikzpicture}

\vspace{-0.3cm}
\caption{The illustration of homomorphic encryption. 
%The full homomorphic encryption is applicable to any degree-bounded polynomial function $f(\cdot)$, without decrypting them first. 
The full homomorphic encryption is applicable to any arbitrary function $f(\cdot)$. 
For speech recognition, the $\mathbf{x}$ can be the input features and $f(\cdot)$ can be the acoustic models. The main challenge is how to run everything efficiently.
%are 1) how to build the acoustic models using polynomial functions; 2) how to run everything efficiently. 
}
\label{fig:HE}\vspace{-0.2cm}
\end{figure}

One way to preserve the privacy of data is to use encryption. Traditional encryption
lock data down in a way that makes it impossible to use, or compute on, unless we decrypt it.
The homomorphic encryption (HE) allows people to use data in computations even while that data are still encrypted \cite{aslett2015review, fan2012somewhat} as illustrated in Figure \ref{fig:HE}.
The encryption is called ``homomorphic" because the transformation has the same effect on both the unencrypted and encrypted data. For example, suppose an encryption algorithm requires multiplying input numbers by $10$ and the decryption requires dividing them by $10$. This encryption is homomorphic for simple addition because ``$1+2$" would be encrypted to ``$10+20$", and decrypting the result by dividing it by $10$ which would get to ``$3$", as expected. A standard encryption algorithm, though, might turn the ``$1$" and ``$2$" into a smiley face and a semicolon. Adding such symbols is nonsensical, making computation impossible. 

There are different degrees of homomorphic encryption, sometimes referred to as \textit{fully} homomorphic compared with \textit{partly}. In the previous paragraph, the example is only partly homomorphic because it works only for additions. The well known RSA algorithm \cite{RSA1978} is also  partly homomorphic. The fully homomorphic encryption allows for any arbitrary function $f(\cdot)$ to be performed on the encrypted data \cite{gentry2009fully},
%number of addition and multiplication operations
\begin{equation}\label{eq:HE}
     %{\mathbf E}^{-1}_k \left[ f({\mathbf E}_k (\x)) \right] \equiv  f({\x})  
     {\enc}^{-1}_k \left[ f({\enc}_k (\x)) \right] \equiv  f({\x})  
\end{equation}
where ${\enc}_k$ and ${\enc}^{-1}_k$ are homomorphic encryption and decryption 
respectively. The subscript $k$ denotes the secret key. In theory, we could treat the acoustic models such as DNNs as the arbitrary function $f(\cdot)$ and apply the homomorphic encryption.
In practice, high degree polynomial function $f(\cdot)$ requires the use of large parameters, which results in larger encrypted messages and slower computation \cite{can-homomorphic-encryption-be-practical-2}. Therefore, for efficiency, it is desired to restrict the acoustic model to degree-bounded polynomials, which only includes additions and multiplications. Thus, to make DNNs compatible with homomorphic encryption some modifications are needed. Details are discussed in Section \ref{sec:PolyNet}.

Another limitation of homomorphic encryption is that it does not support floating-point numbers in practice. We have to use fixed-point real numbers and convert them to integers using the encoding approach in \cite{CryptoNets} for efficiency. We use $4$-$8$ bits of precision on the weights of the network and inputs. To compensate the performance loss caused by low-bit quantization, a special training algorithm is proposed in Section~\ref{sec:PolyNet}. 
%If we train the network in the traditional way with $32$-bit floats and quantize the weights into $8$-bits precision for decoding, the performance will drop substantially. Therefore the retraining with quantization is required (see Section \ref{sec:PolyNet} for details). Empirically, we observed performance degradation $16$-bit

The detailed description of encryption scheme, including the key generation algorithm, encryption and decryption algorithms, are beyond the scope of this paper and can be found in \cite{dowlin2017manual}. In the experiments, we use the implementation of  \textit{Simple Encrypted Arithmetic Library (SEAL)} \cite{sealcrypto} by Microsoft.\footnote{The SEAL library is available at \url{http://sealcrypto.org/}.}

%%%%%%%%%%%%%%%%%%%%%%%%%%%%%%%%%%
\begin{figure}[t]
  \centerline{\includegraphics[width=9cm]{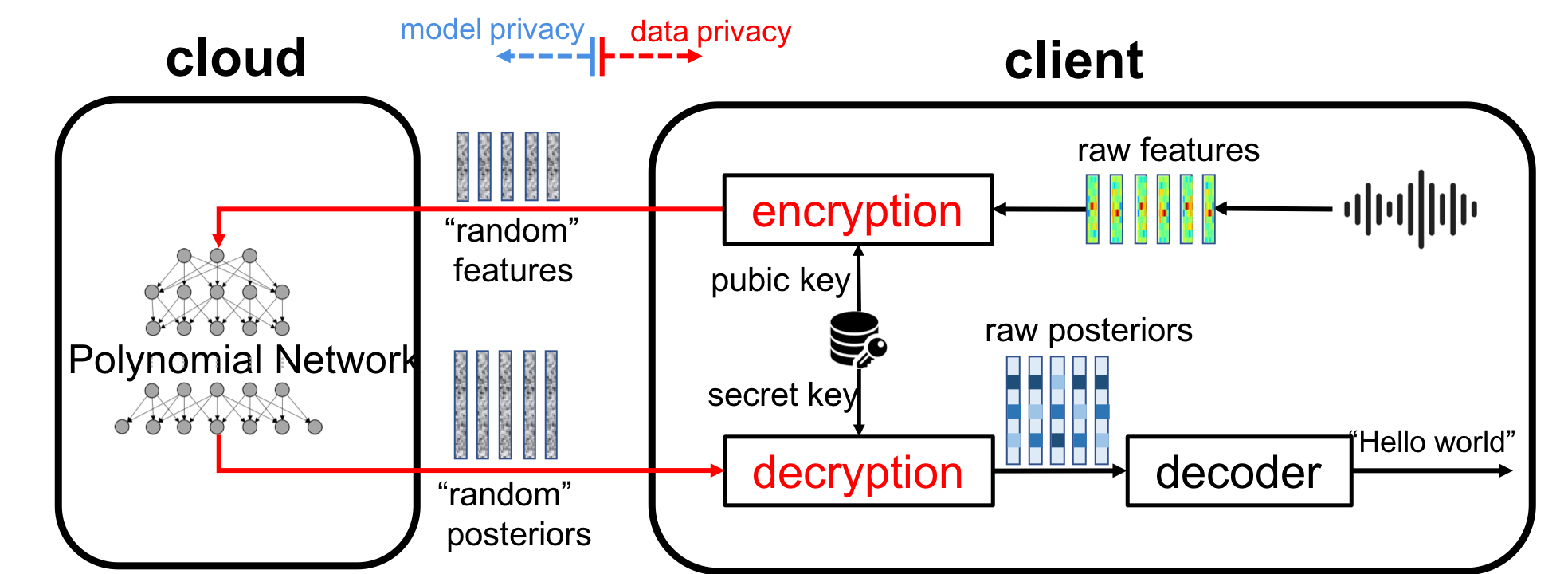}}  \vspace{-0.1cm}
\caption{The framework of encrypted speech recognition. The deep polynomial network is trained on unencrypted data with 8-bit quantization. During decoding the network is operated on encrypted space without knowing secret key.} \vspace{-0.2cm}
\label{fig:framework}
\end{figure}

%%%%%%%%%%%%%%%%%%%%%%%%%%%%%%%%%%
\subsection{Decoding Framework}
\label{ssec:decoding}

In this section we propose a practical framework that enables clients and server to collaboratively decode the speech while satisfying their privacy constraints. During the entire process, the client has no access to the server model and the server has no idea about the input speech and recognition results as illustrated in Figure \ref{fig:framework}. 
First, the clients extract features from the audio in local. Second, the features are encrypted and sent to the cloud. Third, the acoustic models on the cloud evaluate the encrypted features and return the frame-level posteriors in encrypted form. Since the server does not have the secret key, it can not decrypt the posteriors nor decode the utterance. Finally, the client decrypt the frame-level scores and using (personal) language model to decode the utterance in local. These procedures are summarized in Table \ref{tbl:protocol}.

%\noindent 
\vspace{-0.2cm}
\subsubsection{Why not run everything on local} \vspace{-0.15cm}
In theory we could keep everything on local to maintain the privacy. In practice, 
the state-of-art acoustic models in production could be very large (for example up to 1 GB) and the computation can be very intensive. Some models even require specific hardware in order to decode speech in real time. 
It is impossible to run these models on local devices in real time. In addition, the acoustic models in production usually got updated fairly often. It is much easier to deploy the new model if it is run on cloud. Most importantly, running everything on local may divulge the model and decoder to potential hackers.

\vspace{-0.15cm}
\subsubsection{Why not run everything on cloud} \vspace{-0.15cm}
For efficiency we would like to move the decoder and language model to the cloud as well, so that the client only need to do the decryption to get the recognition results. 
Some recent work shown that it is possible to apply the Viterbi search operation over  homomorphic encrypted data \cite{pathak2013privacy, aliasgari2017secure}, however, 
all these papers assume that no pruning is applied while searching in encrypted domain. In practice, speech recognition is never preformed in this manner. On the other hand, pruning 
may restrict the hypothesis set and reveal information about the recognition output. How to prune in encrypted domain and hide this information is still an open problem.

\begin{table}[thb]               % PUT here for align
\begin{center}
\includegraphics[width=8cm]{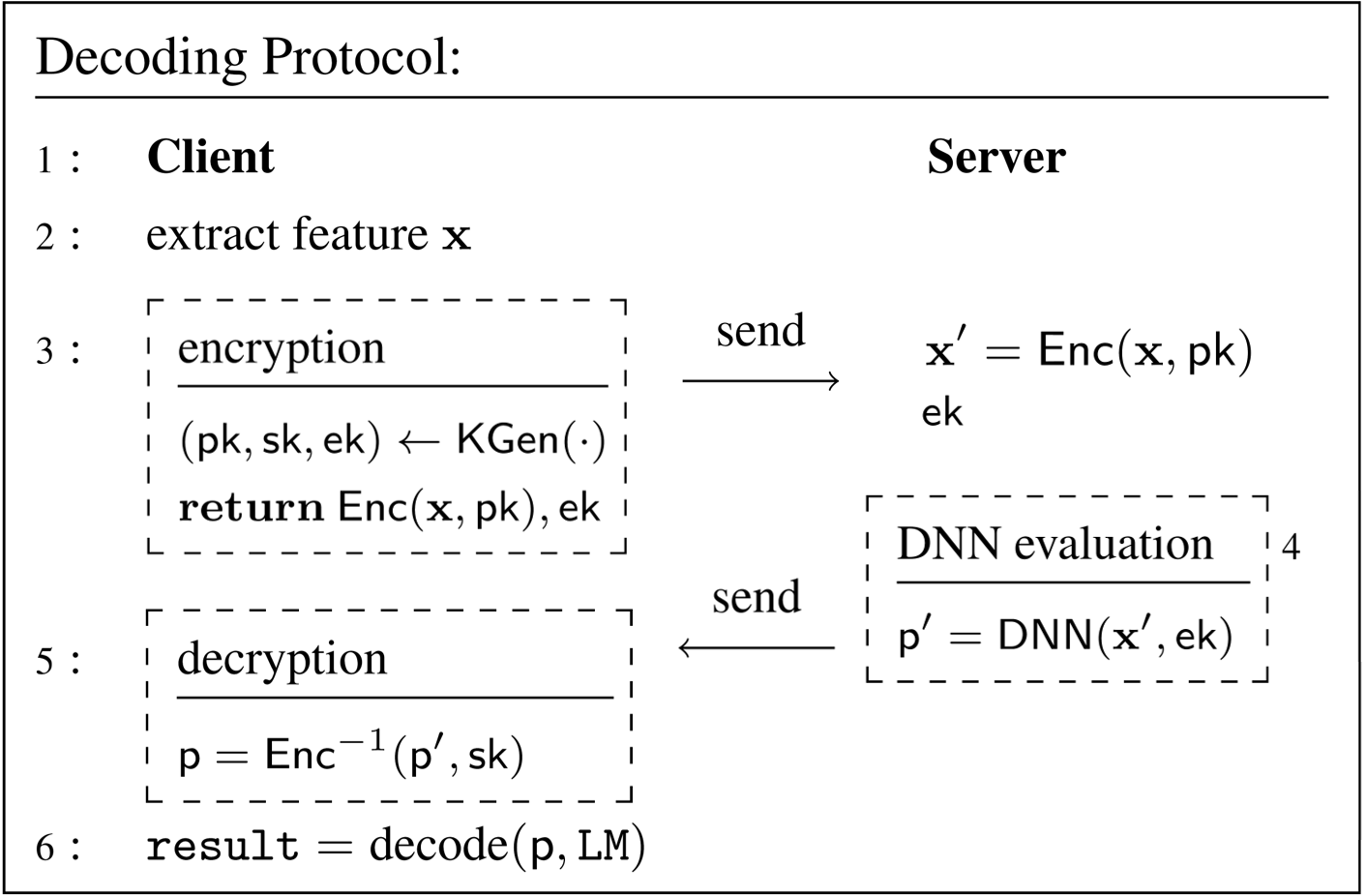}
\end{center}
\vspace{-0.5cm} \caption{\label{tbl:protocol} The decoding protocol for encrypted speech recognition. $(\pk, \sk)$ denotes for a public/secret key pair, $\mathsf{ek}$ is the evaluation key, and $\pp$ is the frame-level posteriors. } \vspace{-0.2cm}
\end{table}

% \pagebreak
% \begin{center}
% \fbox{%
% \procedure[width=8.0cm]{Decoding Protocol: }{%
% \pcln  \textbf{Client}                      \< \< \textbf{Server} \< \\
% \pcln   \text{extract feature } {\mathbf x}  \< \<              \< \\
% \pcln           \begin{subprocedure}%
%               \dbox{\procedure[width=2.8cm]{encryption}{%
%                   (\pk,\sk, \mathsf{ek}) \leftarrow \kgen (\cdot)   \\
%                     \pcreturn \enc({\mathbf x}, \pk), \mathsf{ek} }}
%                 \end{subprocedure} %
%                   \< \sendmessageright*[1cm]{\text{send}} \< {\mathbf x}' = \enc({\mathbf x}, \pk) \<   \\
%  \< \sendmessageleft*[1cm]{\text{send}} \< 
%                 \begin{subprocedure}%
%               \dbox{\procedure[width=2.3cm]{DNN evaluation}{%
%                   \pp' = {\mathsf{DNN}}( {\mathbf x}', \mathsf{ek} )  }}
%                 \end{subprocedure} %
%                 \<  \pclnr \\
% \vspace{-1cm}
%  \pcln \begin{subprocedure}%
%               \dbox{\procedure[width=2.8cm]{decryption}{%
%                   \pp = \enc^{-1}(\pp',\sk) }}
%                 \end{subprocedure} %
%                  \< \< \< \\
%  \pcln {\tt result} = \text{decode}(\pp, \tt LM) \< \< \<
%  }
% }\label{alg:decode}
% \end{center}

\section{Deep Polynomial Networks}
\label{sec:PolyNet}
\vspace{-0.1cm}

In this section we propose a special form of DNNs as acoustic models to make predictions over the encrypted input features. In Section \ref{ssec:HE} we conclude that certain polynomial functions can be computed over encrypted data given that their degree is not too large. %, in the context of homomorphic encryption, 
However, some operations in neural networks are not polynomials, such as sigmoids and ReLU activation functions \cite{maas2013rectifier} and max pooling \cite{krizhevsky2012imagenet}. Here we listed some of common operations in DNNs (including CNNs) and their approximations in polynomials.

\begin{description} %[align=right]
\item [Dense Layer]  This is just linear multiplication and addition. It can be directly implemented for homomorphic encryption without approximation. Note the weights $\mathbf W$ in dense layer are fixed and not encrypted during decoding. Given encrypted inputs $\enc(\mathbf x)$, a naive way to compute dense layers is to first encrypt the weights and then perform the multiplication in encrypted domain $\enc(\mathbf W)^{\sf T}\enc(\mathbf x)$, so that after decryption we can get the exact value of ${\mathbf W}^{\sf T} \mathbf x$. However, this process is computationally intensive and not necessary. Instead, we use a more efficient plain operation ${\mathbf W}^{\sf T}\enc(\mathbf x)$. Because of this, during decryption, we need to keep in mind that the random noise bias in encryption scheme has been scaled by ${\mathbf W}$. 

% \begin{figure}[h]
% \floatbox[{\capbeside\thisfloatsetup{capbesideposition={right,top},capbesidewidth=4cm}}]{figure}[\FBwidth]
% {\caption{The polynomial approximation of Sigmoid and ReLU.}\label{fig:ReLU}}
% {\includegraphics[width=4cm]{ReLU.png}}
% \end{figure}

\item [Batch Norm] This is also multiplications and additions \cite{ioffe2015batch} $ \mathsf{BN}(\mathbf x) = \bm\gamma \frac{{\mathbf x}- \bm\mu}{\bm\sigma}+\bm\beta$, which can be directly implemented for homomorphic encryption without approximation. Note since $\bm \mu, \bm \sigma, \bm\gamma$ and $ \bm\beta$ are fixed during decoding, there is no need to encrypt these parameters or to compute batch norm explicitly. Instead, we merge these parameters to the preceding dense layer, which results to new weights $\mathbf W^{\tt new} = {\tt diag}(\frac{ \bm\gamma}{\bm \sigma}) \mathbf W $ and bias $\bm b^{\tt new} = \bm b + \mathbf W^{\sf T} \bm\beta - \mathbf W^{\sf T} \frac{\bm\mu .\bm\gamma }{\bm \sigma}$ for the preceding layer.

\item [ReLU] This activation function $z \mapsto \max(0, z)$ can be approximated with $\pp(z):=z^2$. Note there is a theoretical study of the problem of learning neural networks with polynomial activation functions in \cite{livni2014computational}, where the above square function was also used to replace ReLU.

\item [Sigmoid] This activation function $z \mapsto \frac{1}{1+ e^{-z} }$ can be approximated with low-degree polynomials $\pp(z) := \frac{1}{2} + \frac{1}{4} z - \frac{1}{48} z^3$.

\item [Convolution] The operation is essentially a dot product of the weight vector (kernel) and the vector of feeding layer outputs. Therefore it is compatible with homomorphic encryption. 

\item [Max Pooling] This operation cannot be computed directly as it is non-polynomial. However, it can be approximated using $\max (z_1, \dots, z_n) = \lim_{d \rightarrow \infty} (\sum_{i=1}^n z_i^d )^{1/d}$. For efficiency, we use $d=1$ which leads to a scaled mean pooling. 

\end{description}

Using the above approximations all the layers and operations in the traditional DNNs and CNNs becomes polynomial. We named the resulting models as deep polynomial networks.

% \begin{itemize}[leftmargin=*,label=\raisebox{0.25ex}{\tiny$\bullet$}]
% \item Dense Layer. This is linear multiplications and additions. It can be directly implemented for homomorphic encryption without approximation. Note the weights in dense layers are not encrypted. 
% Given the encrypt inputs, a naive way to compute the network output is to first encrypt the weights and then perform the addition or multiplication operation in encrypted domain. However, this process is both computationally intensive and not necessary. We can use the more efficient plain operation when compute the multiplications between the fixed weights and the outputs of active functions. 
% \end{itemize}

\SetInd{0.1em}{0.5em}
\setlength{\textfloatsep}{6pt}% Remove \textfloatsep
\begin{algorithm}[t] 
\DontPrintSemicolon
\KwData{unencrypted training data.}
\KwResult{quantized deep polynomial network.}
convert pretrained DNN to DPN: ${\bf W}_{\textcolor{black}{\tt DPN}} \leftarrow  {\bf W}_{\textcolor{black}{\tt DNN}}$ \;
\For{each minibatch}{
\nonl    update ${\bf W}_{\textcolor{black}{\tt DPN}}$ in floating point
}
\For{each layer $[l]$}{
\nonl      optimize codebook: $\textcolor{blue}{\bm c}^{[l]}_{1:256} \leftarrow {\sf{distribution}}( {\bf W}^{[l]}_{\textcolor{black}{\tt DPN}})$ \;
\nonl      quantization: ${\textcolor{blue}{\Tilde{\bf W}}}^{[l]}_{\tt DPN} \leftarrow {\sf{Quant}}( {\bf W}^{[l]}_{\tt DPN}, {\textcolor{blue}{\bm c}}^{[l]}_{1:256} ) $\;
}
\For{each minibatch}{
\nonl  update ${\textcolor{blue}{\Tilde{\bf W}}}_{\tt DPN}$ with on-the-fly quantization based on ${\textcolor{blue}{\bm c}}$
}
\caption{Quantized training for DPN \label{alg:train}} 
\end{algorithm} %

\vspace{-0.2cm}
\subsection{Training with Quantization} 
\vspace{-0.1cm}

In Section \ref{ssec:HE} we discussed that the homomorphic encryption in practice only supports fixed-precision real numbers or integers.\footnote{The SEAL library can encode the fixed-precision real numbers into integers for efficient encryption during decoding \cite{dowlin2017manual}.} However the state-of-the-art DNNs/CNNs are typically trained on GPUs with $32$-bit floating point. Applying low-bit fixed-point quantization directly to the model parameters during decoding could cause substantial performance drop. Here a quantized retraining algorithm is proposed. In Section \ref{sec:Exp} We show that the DNNs, CNNs and DPNs can all be trained using $8$-bit quantization with almost no WER increase.

 %Besides to accommodate the HE, there are other benefits for model quantization. 1) Arithmetic with lower bit is faster. 2) Lighter models are easier to fit into hardware that has very limited on-chip memory such as FPGAs. 3) Lower bit-widths allows us to squeeze more data into the same cache/register. This means we could reduce how often we access data from RAM, which can save a lot of power consumption.

\begin{figure}[t]

\tikzset{every picture/.style={line width=0.75pt}} %set default line width to 0.75pt        

\begin{tikzpicture}[x=0.75pt,y=0.75pt,yscale=-.63,xscale=.63]

%Shape: Rectangle [id:dp4434875262091752] 
\draw  [color={rgb, 255:red, 255; green, 255; blue, 255 }  ,draw opacity=1 ][fill={rgb, 255:red, 128; green, 128; blue, 128 }  ,fill opacity=1 ] (443,74) -- (448,74) -- (448,159) -- (443,159) -- cycle ;
%Shape: Rectangle [id:dp8759782541695786] 
\draw  [color={rgb, 255:red, 255; green, 255; blue, 255 }  ,draw opacity=1 ][fill={rgb, 255:red, 128; green, 128; blue, 128 }  ,fill opacity=1 ] (448,69) -- (453,69) -- (453,159) -- (448,159) -- cycle ;
%Shape: Rectangle [id:dp5335219781261501] 
\draw  [color={rgb, 255:red, 255; green, 255; blue, 255 }  ,draw opacity=1 ][fill={rgb, 255:red, 128; green, 128; blue, 128 }  ,fill opacity=1 ] (438,79) -- (443,79) -- (443,159) -- (438,159) -- cycle ;
%Shape: Rectangle [id:dp8689590860790392] 
\draw  [color={rgb, 255:red, 255; green, 255; blue, 255 }  ,draw opacity=1 ][fill={rgb, 255:red, 128; green, 128; blue, 128 }  ,fill opacity=1 ] (453,69) -- (458,69) -- (458,159) -- (453,159) -- cycle ;
%Shape: Rectangle [id:dp11350264433418211] 
\draw  [color={rgb, 255:red, 255; green, 255; blue, 255 }  ,draw opacity=1 ][fill={rgb, 255:red, 128; green, 128; blue, 128 }  ,fill opacity=1 ] (458,64) -- (463,64) -- (463,159) -- (458,159) -- cycle ;
%Shape: Rectangle [id:dp45588345611377623] 
\draw  [color={rgb, 255:red, 255; green, 255; blue, 255 }  ,draw opacity=1 ][fill={rgb, 255:red, 128; green, 128; blue, 128 }  ,fill opacity=1 ] (468,84) -- (473,84) -- (473,159) -- (468,159) -- cycle ;
%Shape: Rectangle [id:dp25459882987644133] 
\draw  [color={rgb, 255:red, 255; green, 255; blue, 255 }  ,draw opacity=1 ][fill={rgb, 255:red, 128; green, 128; blue, 128 }  ,fill opacity=1 ] (463,79) -- (468,79) -- (468,159) -- (463,159) -- cycle ;
%Shape: Rectangle [id:dp4917259827392084] 
\draw  [color={rgb, 255:red, 255; green, 255; blue, 255 }  ,draw opacity=1 ][fill={rgb, 255:red, 128; green, 128; blue, 128 }  ,fill opacity=1 ] (433,89) -- (438,89) -- (438,159) -- (433,159) -- cycle ;
%Shape: Rectangle [id:dp8869130715214908] 
\draw  [color={rgb, 255:red, 255; green, 255; blue, 255 }  ,draw opacity=1 ][fill={rgb, 255:red, 128; green, 128; blue, 128 }  ,fill opacity=1 ] (473,94) -- (478,94) -- (478,159) -- (473,159) -- cycle ;
%Shape: Rectangle [id:dp06252693752540128] 
\draw  [color={rgb, 255:red, 255; green, 255; blue, 255 }  ,draw opacity=1 ][fill={rgb, 255:red, 128; green, 128; blue, 128 }  ,fill opacity=1 ] (478,104) -- (483,104) -- (483,159) -- (478,159) -- cycle ;
%Shape: Rectangle [id:dp7861021347161399] 
\draw  [color={rgb, 255:red, 255; green, 255; blue, 255 }  ,draw opacity=1 ][fill={rgb, 255:red, 128; green, 128; blue, 128 }  ,fill opacity=1 ] (483,114) -- (488,114) -- (488,159) -- (483,159) -- cycle ;
%Shape: Rectangle [id:dp7117184972396028] 
\draw  [color={rgb, 255:red, 255; green, 255; blue, 255 }  ,draw opacity=1 ][fill={rgb, 255:red, 128; green, 128; blue, 128 }  ,fill opacity=1 ] (488,124) -- (493,124) -- (493,159) -- (488,159) -- cycle ;
%Shape: Rectangle [id:dp24677502534314855] 
\draw  [color={rgb, 255:red, 255; green, 255; blue, 255 }  ,draw opacity=1 ][fill={rgb, 255:red, 128; green, 128; blue, 128 }  ,fill opacity=1 ] (493,134) -- (498,134) -- (498,159) -- (493,159) -- cycle ;
%Shape: Rectangle [id:dp6935446912403744] 
\draw  [color={rgb, 255:red, 255; green, 255; blue, 255 }  ,draw opacity=1 ][fill={rgb, 255:red, 128; green, 128; blue, 128 }  ,fill opacity=1 ] (498,144) -- (503,144) -- (503,159) -- (498,159) -- cycle ;
%Shape: Rectangle [id:dp42028979359646057] 
\draw  [color={rgb, 255:red, 255; green, 255; blue, 255 }  ,draw opacity=1 ][fill={rgb, 255:red, 128; green, 128; blue, 128 }  ,fill opacity=1 ] (503,144) -- (508,144) -- (508,159) -- (503,159) -- cycle ;
%Shape: Rectangle [id:dp49919307874659513] 
\draw  [color={rgb, 255:red, 255; green, 255; blue, 255 }  ,draw opacity=1 ][fill={rgb, 255:red, 128; green, 128; blue, 128 }  ,fill opacity=1 ] (508,149) -- (513,149) -- (513,159) -- (508,159) -- cycle ;
%Shape: Rectangle [id:dp6792651717234593] 
\draw  [color={rgb, 255:red, 255; green, 255; blue, 255 }  ,draw opacity=1 ][fill={rgb, 255:red, 128; green, 128; blue, 128 }  ,fill opacity=1 ] (513,154) -- (518,154) -- (518,159) -- (513,159) -- cycle ;
%Shape: Rectangle [id:dp14174282537861427] 
\draw  [color={rgb, 255:red, 255; green, 255; blue, 255 }  ,draw opacity=1 ][fill={rgb, 255:red, 128; green, 128; blue, 128 }  ,fill opacity=1 ] (428,99) -- (433,99) -- (433,159) -- (428,159) -- cycle ;
%Shape: Rectangle [id:dp4138534863211146] 
\draw  [color={rgb, 255:red, 255; green, 255; blue, 255 }  ,draw opacity=1 ][fill={rgb, 255:red, 128; green, 128; blue, 128 }  ,fill opacity=1 ] (423,104) -- (428,104) -- (428,159) -- (423,159) -- cycle ;
%Shape: Rectangle [id:dp35294828394704103] 
\draw  [color={rgb, 255:red, 255; green, 255; blue, 255 }  ,draw opacity=1 ][fill={rgb, 255:red, 128; green, 128; blue, 128 }  ,fill opacity=1 ] (418,119) -- (423,119) -- (423,159) -- (418,159) -- cycle ;
%Shape: Rectangle [id:dp8558944930275562] 
\draw  [color={rgb, 255:red, 255; green, 255; blue, 255 }  ,draw opacity=1 ][fill={rgb, 255:red, 128; green, 128; blue, 128 }  ,fill opacity=1 ] (413,129) -- (418,129) -- (418,159) -- (413,159) -- cycle ;
%Shape: Rectangle [id:dp9515208679410411] 
\draw  [color={rgb, 255:red, 255; green, 255; blue, 255 }  ,draw opacity=1 ][fill={rgb, 255:red, 128; green, 128; blue, 128 }  ,fill opacity=1 ] (408,139) -- (413,139) -- (413,159) -- (408,159) -- cycle ;
%Shape: Rectangle [id:dp9472243594136989] 
\draw  [color={rgb, 255:red, 255; green, 255; blue, 255 }  ,draw opacity=1 ][fill={rgb, 255:red, 128; green, 128; blue, 128 }  ,fill opacity=1 ] (403,144) -- (408,144) -- (408,159) -- (403,159) -- cycle ;
%Shape: Rectangle [id:dp7355570456311114] 
\draw  [color={rgb, 255:red, 255; green, 255; blue, 255 }  ,draw opacity=1 ][fill={rgb, 255:red, 128; green, 128; blue, 128 }  ,fill opacity=1 ] (393,149) -- (398,149) -- (398,159) -- (393,159) -- cycle ;
%Shape: Rectangle [id:dp2757509960582055] 
\draw  [color={rgb, 255:red, 255; green, 255; blue, 255 }  ,draw opacity=1 ][fill={rgb, 255:red, 128; green, 128; blue, 128 }  ,fill opacity=1 ] (398,144) -- (403,144) -- (403,159) -- (398,159) -- cycle ;
%Shape: Rectangle [id:dp5166507681728225] 
\draw  [color={rgb, 255:red, 255; green, 255; blue, 255 }  ,draw opacity=1 ][fill={rgb, 255:red, 128; green, 128; blue, 128 }  ,fill opacity=1 ] (388,149) -- (393,149) -- (393,159) -- (388,159) -- cycle ;
%Shape: Rectangle [id:dp8237021650571752] 
\draw  [color={rgb, 255:red, 255; green, 255; blue, 255 }  ,draw opacity=1 ][fill={rgb, 255:red, 128; green, 128; blue, 128 }  ,fill opacity=1 ] (383,154) -- (388,154) -- (388,159) -- (383,159) -- cycle ;
%Straight Lines [id:da508062798647887] 
\draw    (116.17,219.1) -- (116.17,205.1) ;
\draw [shift={(116.17,203.1)}, rotate = 450] [color={rgb, 255:red, 0; green, 0; blue, 0 }  ][line width=0.75]    (10.93,-3.29) .. controls (6.95,-1.4) and (3.31,-0.3) .. (0,0) .. controls (3.31,0.3) and (6.95,1.4) .. (10.93,3.29)   ;

%Rounded Rect [id:dp09161629457378728] 
\draw  [fill={rgb, 255:red, 74; green, 74; blue, 74 }  ,fill opacity=1 ] (212.5,224.58) .. controls (212.5,221.83) and (214.73,219.6) .. (217.48,219.6) -- (289.52,219.6) .. controls (292.27,219.6) and (294.5,221.83) .. (294.5,224.58) -- (294.5,239.52) .. controls (294.5,242.27) and (292.27,244.5) .. (289.52,244.5) -- (217.48,244.5) .. controls (214.73,244.5) and (212.5,242.27) .. (212.5,239.52) -- cycle ;
%Straight Lines [id:da7499889507210773] 
\draw    (254.17,221.1) -- (254.17,203.1) ;
\draw [shift={(254.17,201.1)}, rotate = 450] [color={rgb, 255:red, 0; green, 0; blue, 0 }  ][line width=0.75]    (10.93,-3.29) .. controls (6.95,-1.4) and (3.31,-0.3) .. (0,0) .. controls (3.31,0.3) and (6.95,1.4) .. (10.93,3.29)   ;

%Curve Lines [id:da7791864442892609] 
\draw    (116.17,167.1) .. controls (124.05,146.42) and (133.87,147.07) .. (170.48,148.06) ;
\draw [shift={(172.17,148.1)}, rotate = 181.51] [color={rgb, 255:red, 0; green, 0; blue, 0 }  ][line width=0.75]    (10.93,-3.29) .. controls (6.95,-1.4) and (3.31,-0.3) .. (0,0) .. controls (3.31,0.3) and (6.95,1.4) .. (10.93,3.29)   ;

%Curve Lines [id:da8330388055178334] 
\draw    (252.17,166.1) .. controls (242.32,143.45) and (228.59,147.96) .. (193.77,148.1) ;
\draw [shift={(192.17,148.1)}, rotate = 360] [color={rgb, 255:red, 0; green, 0; blue, 0 }  ][line width=0.75]    (10.93,-3.29) .. controls (6.95,-1.4) and (3.31,-0.3) .. (0,0) .. controls (3.31,0.3) and (6.95,1.4) .. (10.93,3.29)   ;

\draw   (176.61,144.39) .. controls (179.59,141.41) and (184.45,141.45) .. (187.47,144.47) .. controls (190.49,147.49) and (190.53,152.35) .. (187.55,155.32) .. controls (184.58,158.3) and (179.72,158.26) .. (176.7,155.24) .. controls (173.68,152.22) and (173.64,147.36) .. (176.61,144.39) -- cycle ; \draw   (176.61,144.39) -- (187.55,155.32) ; \draw   (187.47,144.47) -- (176.7,155.24) ;
%Straight Lines [id:da8073862299400109] 
\draw    (182.47,102.33) -- (182.47,84.33) ;
\draw [shift={(182.47,82.33)}, rotate = 450] [color={rgb, 255:red, 0; green, 0; blue, 0 }  ][line width=0.75]    (10.93,-3.29) .. controls (6.95,-1.4) and (3.31,-0.3) .. (0,0) .. controls (3.31,0.3) and (6.95,1.4) .. (10.93,3.29)   ;

%Straight Lines [id:da41767486718618396] 
\draw    (182.17,141.1) -- (182.43,127.33) ;
\draw [shift={(182.47,125.33)}, rotate = 451.09] [color={rgb, 255:red, 0; green, 0; blue, 0 }  ][line width=0.75]    (10.93,-3.29) .. controls (6.95,-1.4) and (3.31,-0.3) .. (0,0) .. controls (3.31,0.3) and (6.95,1.4) .. (10.93,3.29)   ;

%Straight Lines [id:da05913044218198449] 
\draw [color={rgb, 255:red, 128; green, 128; blue, 128 }  ,draw opacity=1 ]   (339.47,163.43) -- (577.47,163.04) (344.46,160.92) -- (344.47,165.92)(349.46,160.91) -- (349.47,165.91)(354.46,160.91) -- (354.47,165.91)(359.46,160.9) -- (359.47,165.9)(364.46,160.89) -- (364.47,165.89)(369.46,160.88) -- (369.47,165.88)(374.46,160.87) -- (374.47,165.87)(379.46,160.86) -- (379.47,165.86)(384.46,160.86) -- (384.47,165.86)(389.46,160.85) -- (389.47,165.85)(394.46,160.84) -- (394.47,165.84)(399.46,160.83) -- (399.47,165.83)(404.46,160.82) -- (404.47,165.82)(409.46,160.81) -- (409.47,165.81)(414.46,160.81) -- (414.47,165.81)(419.46,160.8) -- (419.47,165.8)(424.46,160.79) -- (424.47,165.79)(429.46,160.78) -- (429.47,165.78)(434.46,160.77) -- (434.47,165.77)(439.46,160.76) -- (439.47,165.76)(444.46,160.76) -- (444.47,165.76)(449.46,160.75) -- (449.47,165.75)(454.46,160.74) -- (454.47,165.74)(459.46,160.73) -- (459.47,165.73)(464.46,160.72) -- (464.47,165.72)(469.46,160.72) -- (469.47,165.72)(474.46,160.71) -- (474.47,165.71)(479.46,160.7) -- (479.47,165.7)(484.46,160.69) -- (484.47,165.69)(489.46,160.68) -- (489.47,165.68)(494.46,160.67) -- (494.47,165.67)(499.46,160.67) -- (499.47,165.67)(504.46,160.66) -- (504.47,165.66)(509.46,160.65) -- (509.47,165.65)(514.46,160.64) -- (514.47,165.64)(519.46,160.63) -- (519.47,165.63)(524.46,160.62) -- (524.47,165.62)(529.46,160.62) -- (529.47,165.62)(534.46,160.61) -- (534.47,165.61)(539.46,160.6) -- (539.47,165.6)(544.46,160.59) -- (544.47,165.59)(549.46,160.58) -- (549.47,165.58)(554.46,160.57) -- (554.47,165.57)(559.46,160.57) -- (559.47,165.57)(564.46,160.56) -- (564.47,165.56)(569.46,160.55) -- (569.47,165.55)(574.46,160.54) -- (574.47,165.54) ;
\draw [shift={(579.47,163.03)}, rotate = 539.9100000000001] [fill={rgb, 255:red, 128; green, 128; blue, 128 }  ,fill opacity=1 ][line width=0.75]  [draw opacity=0] (8.93,-4.29) -- (0,0) -- (8.93,4.29) -- cycle    ;
\draw [shift={(337.47,163.43)}, rotate = 359.91] [fill={rgb, 255:red, 128; green, 128; blue, 128 }  ,fill opacity=1 ][line width=0.75]  [draw opacity=0] (8.93,-4.29) -- (0,0) -- (8.93,4.29) -- cycle    ;
%Straight Lines [id:da3381886610245156] 
\draw [color={rgb, 255:red, 128; green, 128; blue, 128 }  ,draw opacity=1 ]   (339.47,189.43) -- (577.47,189.04) (359.46,187.4) -- (359.47,191.4)(379.46,187.36) -- (379.47,191.36)(399.46,187.33) -- (399.47,191.33)(419.46,187.3) -- (419.47,191.3)(439.46,187.26) -- (439.47,191.26)(459.46,187.23) -- (459.47,191.23)(479.46,187.2) -- (479.47,191.2)(499.46,187.17) -- (499.47,191.17)(519.46,187.13) -- (519.47,191.13)(539.46,187.1) -- (539.47,191.1)(559.46,187.07) -- (559.47,191.07) ;
\draw [shift={(579.47,189.03)}, rotate = 539.9100000000001] [fill={rgb, 255:red, 128; green, 128; blue, 128 }  ,fill opacity=1 ][line width=0.75]  [draw opacity=0] (8.93,-4.29) -- (0,0) -- (8.93,4.29) -- cycle    ;
\draw [shift={(337.47,189.43)}, rotate = 359.91] [fill={rgb, 255:red, 128; green, 128; blue, 128 }  ,fill opacity=1 ][line width=0.75]  [draw opacity=0] (8.93,-4.29) -- (0,0) -- (8.93,4.29) -- cycle    ;
%Shape: Circle [id:dp9316666812560074] 
\draw  [color={rgb, 255:red, 74; green, 144; blue, 226 }  ,draw opacity=1 ][fill={rgb, 255:red, 74; green, 144; blue, 226 }  ,fill opacity=1 ] (436.57,188.72) .. controls (436.57,187.22) and (437.78,186) .. (439.28,186) .. controls (440.78,186) and (442,187.22) .. (442,188.72) .. controls (442,190.22) and (440.78,191.43) .. (439.28,191.43) .. controls (437.78,191.43) and (436.57,190.22) .. (436.57,188.72) -- cycle ;
%Shape: Circle [id:dp46900994374212546] 
\draw  [color={rgb, 255:red, 74; green, 144; blue, 226 }  ,draw opacity=1 ][fill={rgb, 255:red, 74; green, 144; blue, 226 }  ,fill opacity=1 ] (496.57,188.72) .. controls (496.57,187.22) and (497.78,186) .. (499.28,186) .. controls (500.78,186) and (502,187.22) .. (502,188.72) .. controls (502,190.22) and (500.78,191.43) .. (499.28,191.43) .. controls (497.78,191.43) and (496.57,190.22) .. (496.57,188.72) -- cycle ;
%Shape: Circle [id:dp6466714855973528] 
\draw  [color={rgb, 255:red, 74; green, 144; blue, 226 }  ,draw opacity=1 ][fill={rgb, 255:red, 74; green, 144; blue, 226 }  ,fill opacity=1 ] (456.57,188.72) .. controls (456.57,187.22) and (457.78,186) .. (459.28,186) .. controls (460.78,186) and (462,187.22) .. (462,188.72) .. controls (462,190.22) and (460.78,191.43) .. (459.28,191.43) .. controls (457.78,191.43) and (456.57,190.22) .. (456.57,188.72) -- cycle ;
%Shape: Circle [id:dp2931780877965615] 
\draw  [color={rgb, 255:red, 74; green, 144; blue, 226 }  ,draw opacity=1 ][fill={rgb, 255:red, 74; green, 144; blue, 226 }  ,fill opacity=1 ] (476.57,188.72) .. controls (476.57,187.22) and (477.78,186) .. (479.28,186) .. controls (480.78,186) and (482,187.22) .. (482,188.72) .. controls (482,190.22) and (480.78,191.43) .. (479.28,191.43) .. controls (477.78,191.43) and (476.57,190.22) .. (476.57,188.72) -- cycle ;
%Shape: Circle [id:dp78742708195132] 
\draw  [color={rgb, 255:red, 74; green, 144; blue, 226 }  ,draw opacity=1 ][fill={rgb, 255:red, 74; green, 144; blue, 226 }  ,fill opacity=1 ] (356.57,188.72) .. controls (356.57,187.22) and (357.78,186) .. (359.28,186) .. controls (360.78,186) and (362,187.22) .. (362,188.72) .. controls (362,190.22) and (360.78,191.43) .. (359.28,191.43) .. controls (357.78,191.43) and (356.57,190.22) .. (356.57,188.72) -- cycle ;
%Shape: Circle [id:dp1623344864311438] 
\draw  [color={rgb, 255:red, 74; green, 144; blue, 226 }  ,draw opacity=1 ][fill={rgb, 255:red, 74; green, 144; blue, 226 }  ,fill opacity=1 ] (416.57,188.72) .. controls (416.57,187.22) and (417.78,186) .. (419.28,186) .. controls (420.78,186) and (422,187.22) .. (422,188.72) .. controls (422,190.22) and (420.78,191.43) .. (419.28,191.43) .. controls (417.78,191.43) and (416.57,190.22) .. (416.57,188.72) -- cycle ;
%Shape: Circle [id:dp2593007680793974] 
\draw  [color={rgb, 255:red, 74; green, 144; blue, 226 }  ,draw opacity=1 ][fill={rgb, 255:red, 74; green, 144; blue, 226 }  ,fill opacity=1 ] (376.57,188.72) .. controls (376.57,187.22) and (377.78,186) .. (379.28,186) .. controls (380.78,186) and (382,187.22) .. (382,188.72) .. controls (382,190.22) and (380.78,191.43) .. (379.28,191.43) .. controls (377.78,191.43) and (376.57,190.22) .. (376.57,188.72) -- cycle ;
%Shape: Circle [id:dp22965581005331404] 
\draw  [color={rgb, 255:red, 74; green, 144; blue, 226 }  ,draw opacity=1 ][fill={rgb, 255:red, 74; green, 144; blue, 226 }  ,fill opacity=1 ] (396.57,188.72) .. controls (396.57,187.22) and (397.78,186) .. (399.28,186) .. controls (400.78,186) and (402,187.22) .. (402,188.72) .. controls (402,190.22) and (400.78,191.43) .. (399.28,191.43) .. controls (397.78,191.43) and (396.57,190.22) .. (396.57,188.72) -- cycle ;
%Shape: Circle [id:dp08547054433785517] 
\draw  [color={rgb, 255:red, 74; green, 144; blue, 226 }  ,draw opacity=1 ][fill={rgb, 255:red, 74; green, 144; blue, 226 }  ,fill opacity=1 ] (558.57,188.72) .. controls (558.57,187.22) and (559.78,186) .. (561.28,186) .. controls (562.78,186) and (564,187.22) .. (564,188.72) .. controls (564,190.22) and (562.78,191.43) .. (561.28,191.43) .. controls (559.78,191.43) and (558.57,190.22) .. (558.57,188.72) -- cycle ;
%Shape: Circle [id:dp5606583501944555] 
\draw  [color={rgb, 255:red, 74; green, 144; blue, 226 }  ,draw opacity=1 ][fill={rgb, 255:red, 74; green, 144; blue, 226 }  ,fill opacity=1 ] (518.57,188.72) .. controls (518.57,187.22) and (519.78,186) .. (521.28,186) .. controls (522.78,186) and (524,187.22) .. (524,188.72) .. controls (524,190.22) and (522.78,191.43) .. (521.28,191.43) .. controls (519.78,191.43) and (518.57,190.22) .. (518.57,188.72) -- cycle ;
%Shape: Circle [id:dp7040432494000382] 
\draw  [color={rgb, 255:red, 74; green, 144; blue, 226 }  ,draw opacity=1 ][fill={rgb, 255:red, 74; green, 144; blue, 226 }  ,fill opacity=1 ] (538.57,188.72) .. controls (538.57,187.22) and (539.78,186) .. (541.28,186) .. controls (542.78,186) and (544,187.22) .. (544,188.72) .. controls (544,190.22) and (542.78,191.43) .. (541.28,191.43) .. controls (539.78,191.43) and (538.57,190.22) .. (538.57,188.72) -- cycle ;
%Straight Lines [id:da10991372444772429] 
\draw [color={rgb, 255:red, 128; green, 128; blue, 128 }  ,draw opacity=1 ][fill={rgb, 255:red, 128; green, 128; blue, 128 }  ,fill opacity=1 ]   (339.57,230.17) -- (577.57,229.78) ;
\draw [shift={(579.57,229.77)}, rotate = 539.9100000000001] [fill={rgb, 255:red, 128; green, 128; blue, 128 }  ,fill opacity=1 ][line width=0.75]  [draw opacity=0] (8.93,-4.29) -- (0,0) -- (8.93,4.29) -- cycle    ;
\draw [shift={(337.57,230.18)}, rotate = 359.91] [fill={rgb, 255:red, 128; green, 128; blue, 128 }  ,fill opacity=1 ][line width=0.75]  [draw opacity=0] (8.93,-4.29) -- (0,0) -- (8.93,4.29) -- cycle    ;
%Shape: Circle [id:dp20990288297386683] 
\draw  [color={rgb, 255:red, 74; green, 144; blue, 226 }  ,draw opacity=1 ][fill={rgb, 255:red, 74; green, 144; blue, 226 }  ,fill opacity=1 ] (431.57,229.97) .. controls (431.57,228.47) and (432.78,227.26) .. (434.28,227.26) .. controls (435.78,227.26) and (437,228.47) .. (437,229.97) .. controls (437,231.48) and (435.78,232.69) .. (434.28,232.69) .. controls (432.78,232.69) and (431.57,231.48) .. (431.57,229.97) -- cycle ;
%Shape: Circle [id:dp8411873014024622] 
\draw  [color={rgb, 255:red, 74; green, 144; blue, 226 }  ,draw opacity=1 ][fill={rgb, 255:red, 74; green, 144; blue, 226 }  ,fill opacity=1 ] (466.57,229.97) .. controls (466.57,228.47) and (467.78,227.26) .. (469.28,227.26) .. controls (470.78,227.26) and (472,228.47) .. (472,229.97) .. controls (472,231.48) and (470.78,232.69) .. (469.28,232.69) .. controls (467.78,232.69) and (466.57,231.48) .. (466.57,229.97) -- cycle ;
%Shape: Circle [id:dp6180956360677813] 
\draw  [color={rgb, 255:red, 74; green, 144; blue, 226 }  ,draw opacity=1 ][fill={rgb, 255:red, 74; green, 144; blue, 226 }  ,fill opacity=1 ] (444.7,229.97) .. controls (444.7,228.47) and (445.92,227.26) .. (447.42,227.26) .. controls (448.92,227.26) and (450.13,228.47) .. (450.13,229.97) .. controls (450.13,231.48) and (448.92,232.69) .. (447.42,232.69) .. controls (445.92,232.69) and (444.7,231.48) .. (444.7,229.97) -- cycle ;
%Shape: Circle [id:dp23701939227304636] 
\draw  [color={rgb, 255:red, 74; green, 144; blue, 226 }  ,draw opacity=1 ][fill={rgb, 255:red, 74; green, 144; blue, 226 }  ,fill opacity=1 ] (461.13,229.97) .. controls (461.13,228.47) and (462.35,227.26) .. (463.85,227.26) .. controls (465.35,227.26) and (466.57,228.47) .. (466.57,229.97) .. controls (466.57,231.48) and (465.35,232.69) .. (463.85,232.69) .. controls (462.35,232.69) and (461.13,231.48) .. (461.13,229.97) -- cycle ;
%Shape: Circle [id:dp6249084976776902] 
\draw  [color={rgb, 255:red, 74; green, 144; blue, 226 }  ,draw opacity=1 ][fill={rgb, 255:red, 74; green, 144; blue, 226 }  ,fill opacity=1 ] (558.57,229.97) .. controls (558.57,228.47) and (559.78,227.26) .. (561.28,227.26) .. controls (562.78,227.26) and (564,228.47) .. (564,229.97) .. controls (564,231.48) and (562.78,232.69) .. (561.28,232.69) .. controls (559.78,232.69) and (558.57,231.48) .. (558.57,229.97) -- cycle ;
%Shape: Circle [id:dp1077058226403329] 
\draw  [color={rgb, 255:red, 74; green, 144; blue, 226 }  ,draw opacity=1 ][fill={rgb, 255:red, 74; green, 144; blue, 226 }  ,fill opacity=1 ] (411.57,229.97) .. controls (411.57,228.47) and (412.78,227.26) .. (414.28,227.26) .. controls (415.78,227.26) and (417,228.47) .. (417,229.97) .. controls (417,231.48) and (415.78,232.69) .. (414.28,232.69) .. controls (412.78,232.69) and (411.57,231.48) .. (411.57,229.97) -- cycle ;
%Shape: Circle [id:dp7682055577446086] 
\draw  [color={rgb, 255:red, 74; green, 144; blue, 226 }  ,draw opacity=1 ][fill={rgb, 255:red, 74; green, 144; blue, 226 }  ,fill opacity=1 ] (356.57,229.97) .. controls (356.57,228.47) and (357.78,227.26) .. (359.28,227.26) .. controls (360.78,227.26) and (362,228.47) .. (362,229.97) .. controls (362,231.48) and (360.78,232.69) .. (359.28,232.69) .. controls (357.78,232.69) and (356.57,231.48) .. (356.57,229.97) -- cycle ;
%Shape: Circle [id:dp6146551589191653] 
\draw  [color={rgb, 255:red, 74; green, 144; blue, 226 }  ,draw opacity=1 ][fill={rgb, 255:red, 74; green, 144; blue, 226 }  ,fill opacity=1 ] (387.57,229.97) .. controls (387.57,228.47) and (388.78,227.26) .. (390.28,227.26) .. controls (391.78,227.26) and (393,228.47) .. (393,229.97) .. controls (393,231.48) and (391.78,232.69) .. (390.28,232.69) .. controls (388.78,232.69) and (387.57,231.48) .. (387.57,229.97) -- cycle ;
%Shape: Circle [id:dp06164143633576957] 
\draw  [color={rgb, 255:red, 74; green, 144; blue, 226 }  ,draw opacity=1 ][fill={rgb, 255:red, 74; green, 144; blue, 226 }  ,fill opacity=1 ] (528.57,229.97) .. controls (528.57,228.47) and (529.78,227.26) .. (531.28,227.26) .. controls (532.78,227.26) and (534,228.47) .. (534,229.97) .. controls (534,231.48) and (532.78,232.69) .. (531.28,232.69) .. controls (529.78,232.69) and (528.57,231.48) .. (528.57,229.97) -- cycle ;
%Shape: Circle [id:dp12118689589894238] 
\draw  [color={rgb, 255:red, 74; green, 144; blue, 226 }  ,draw opacity=1 ][fill={rgb, 255:red, 74; green, 144; blue, 226 }  ,fill opacity=1 ] (476,229.97) .. controls (476,228.47) and (477.22,227.26) .. (478.72,227.26) .. controls (480.22,227.26) and (481.43,228.47) .. (481.43,229.97) .. controls (481.43,231.48) and (480.22,232.69) .. (478.72,232.69) .. controls (477.22,232.69) and (476,231.48) .. (476,229.97) -- cycle ;
%Shape: Circle [id:dp9578580648655537] 
\draw  [color={rgb, 255:red, 74; green, 144; blue, 226 }  ,draw opacity=1 ][fill={rgb, 255:red, 74; green, 144; blue, 226 }  ,fill opacity=1 ] (503.57,229.97) .. controls (503.57,228.47) and (504.78,227.26) .. (506.28,227.26) .. controls (507.78,227.26) and (509,228.47) .. (509,229.97) .. controls (509,231.48) and (507.78,232.69) .. (506.28,232.69) .. controls (504.78,232.69) and (503.57,231.48) .. (503.57,229.97) -- cycle ;
%Shape: Circle [id:dp44355810863577205] 
\draw  [color={rgb, 255:red, 74; green, 144; blue, 226 }  ,draw opacity=1 ][fill={rgb, 255:red, 74; green, 144; blue, 226 }  ,fill opacity=1 ] (455.13,229.97) .. controls (455.13,228.47) and (456.35,227.26) .. (457.85,227.26) .. controls (459.35,227.26) and (460.57,228.47) .. (460.57,229.97) .. controls (460.57,231.48) and (459.35,232.69) .. (457.85,232.69) .. controls (456.35,232.69) and (455.13,231.48) .. (455.13,229.97) -- cycle ;
%Shape: Circle [id:dp08018945901609742] 
\draw  [color={rgb, 255:red, 74; green, 144; blue, 226 }  ,draw opacity=1 ][fill={rgb, 255:red, 74; green, 144; blue, 226 }  ,fill opacity=1 ] (488.57,229.97) .. controls (488.57,228.47) and (489.78,227.26) .. (491.28,227.26) .. controls (492.78,227.26) and (494,228.47) .. (494,229.97) .. controls (494,231.48) and (492.78,232.69) .. (491.28,232.69) .. controls (489.78,232.69) and (488.57,231.48) .. (488.57,229.97) -- cycle ;
%Rounded Rect [id:dp6892947023412294] 
\draw  [draw opacity=0][fill={rgb, 255:red, 74; green, 144; blue, 226 }  ,fill opacity=1 ] (208,174.5) .. controls (208,170.36) and (211.36,167) .. (215.5,167) -- (291,167) .. controls (295.14,167) and (298.5,170.36) .. (298.5,174.5) -- (298.5,197.01) .. controls (298.5,201.16) and (295.14,204.52) .. (291,204.52) -- (215.5,204.52) .. controls (211.36,204.52) and (208,201.16) .. (208,197.01) -- cycle ;
%Rounded Rect [id:dp7700825834810002] 
\draw  [fill={rgb, 255:red, 74; green, 74; blue, 74 }  ,fill opacity=1 ] (150,104.08) .. controls (150,101.33) and (152.23,99.1) .. (154.98,99.1) -- (210.02,99.1) .. controls (212.77,99.1) and (215,101.33) .. (215,104.08) -- (215,119.02) .. controls (215,121.77) and (212.77,124) .. (210.02,124) -- (154.98,124) .. controls (152.23,124) and (150,121.77) .. (150,119.02) -- cycle ;
%Rounded Rect [id:dp12395880302254536] 
\draw  [fill={rgb, 255:red, 74; green, 74; blue, 74 }  ,fill opacity=1 ] (74.5,224.58) .. controls (74.5,221.83) and (76.73,219.6) .. (79.48,219.6) -- (151.52,219.6) .. controls (154.27,219.6) and (156.5,221.83) .. (156.5,224.58) -- (156.5,239.52) .. controls (156.5,242.27) and (154.27,244.5) .. (151.52,244.5) -- (79.48,244.5) .. controls (76.73,244.5) and (74.5,242.27) .. (74.5,239.52) -- cycle ;
%Rounded Rect [id:dp9653349232598668] 
\draw  [draw opacity=0][fill={rgb, 255:red, 74; green, 144; blue, 226 }  ,fill opacity=1 ] (73,173.5) .. controls (73,169.36) and (76.36,166) .. (80.5,166) -- (156,166) .. controls (160.14,166) and (163.5,169.36) .. (163.5,173.5) -- (163.5,196.01) .. controls (163.5,200.16) and (160.14,203.52) .. (156,203.52) -- (80.5,203.52) .. controls (76.36,203.52) and (73,200.16) .. (73,196.01) -- cycle ;
%Rounded Rect [id:dp5400723666895543] 
\draw  [draw opacity=0][fill={rgb, 255:red, 74; green, 144; blue, 226 }  ,fill opacity=1 ] (138,56.5) .. controls (138,52.36) and (141.36,49) .. (145.5,49) -- (221,49) .. controls (225.14,49) and (228.5,52.36) .. (228.5,56.5) -- (228.5,79.01) .. controls (228.5,83.16) and (225.14,86.52) .. (221,86.52) -- (145.5,86.52) .. controls (141.36,86.52) and (138,83.16) .. (138,79.01) -- cycle ;
%Straight Lines [id:da24929484229543686] 
\draw [color={rgb, 255:red, 128; green, 128; blue, 128 }  ,draw opacity=1 ]   (447.82,162) -- (447.5,174.52) -- (439.5,174.52) -- (439.31,189.43) ;
\draw [shift={(439.28,191.43)}, rotate = 270.73] [color={rgb, 255:red, 128; green, 128; blue, 128 }  ,draw opacity=1 ][line width=0.75]    (10.93,-3.29) .. controls (6.95,-1.4) and (3.31,-0.3) .. (0,0) .. controls (3.31,0.3) and (6.95,1.4) .. (10.93,3.29)   ;
\draw [shift={(447.82,162)}, rotate = 91.45] [color={rgb, 255:red, 128; green, 128; blue, 128 }  ,draw opacity=1 ][fill={rgb, 255:red, 128; green, 128; blue, 128 }  ,fill opacity=1 ][line width=0.75]      (0, 0) circle [x radius= 3.35, y radius= 3.35]   ;
%Straight Lines [id:da3822546492475122] 
\draw [color={rgb, 255:red, 128; green, 128; blue, 128 }  ,draw opacity=1 ]   (492,162) -- (492.5,174.52) -- (500.5,174.52) -- (499.61,186.44) ;
\draw [shift={(499.47,188.43)}, rotate = 274.25] [color={rgb, 255:red, 128; green, 128; blue, 128 }  ,draw opacity=1 ][line width=0.75]    (10.93,-3.29) .. controls (6.95,-1.4) and (3.31,-0.3) .. (0,0) .. controls (3.31,0.3) and (6.95,1.4) .. (10.93,3.29)   ;
\draw [shift={(492,162)}, rotate = 87.71] [color={rgb, 255:red, 128; green, 128; blue, 128 }  ,draw opacity=1 ][fill={rgb, 255:red, 128; green, 128; blue, 128 }  ,fill opacity=1 ][line width=0.75]      (0, 0) circle [x radius= 3.35, y radius= 3.35]   ;

% Text Node
\draw (115.5,232.05) node  [align=left] {\textcolor[rgb]{1,1,1}{{\small weights}}};
% Text Node
\draw (253.5,231.55) node [scale=0.9] [align=left] {\textcolor[rgb]{1,1,1}{{\small activations}}};
% Text Node
\draw (182.5,111.55) node  [align=left] {\textcolor[rgb]{1,1,1}{{\small ReLU}}};
% Text Node
\draw (369,170) node  [align=left] {{\footnotesize min}};
% Text Node
\draw (549,170) node  [align=left] {{\footnotesize max}};
% Text Node
\draw (466,202.5) node [scale=1] [align=left] {{\small uniform quantization}};
% Text Node
\draw (468.5,244.5) node [scale=1] [align=left] {{\small Lloyd-max quantization}};
% Text Node
\draw (456.5,52.5) node [scale=1] [align=left] {{\small histgram of weights}};
% Text Node
\draw (253.25,185.76) node  [align=left] {\textcolor[rgb]{1,1,1}{{\scriptsize \textbf{quantizaiton}}}};
% Text Node
\draw (118.25,184.76) node  [align=left] {\textcolor[rgb]{1,1,1}{{\scriptsize \textbf{quantizaiton}}}};
% Text Node
\draw (183.25,67.76) node  [align=left] {\textcolor[rgb]{1,1,1}{{\scriptsize \textbf{quantizaiton}}}};
\vspace{-0.1cm}
\end{tikzpicture}
\vspace{-0.5cm}
\caption{Quantization scheme in training over raw data. } \vspace{-0.1cm}
\label{fig:quant}
\end{figure}
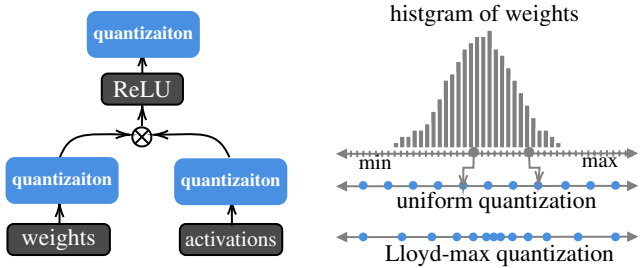

Unlike floating point has an universal standard, the fixed-point numbers are domain specific. Each task has to design its own scheme for fixed-point quantization \cite{lin2016fixed, han2015deep}. Our quantization scheme has three features. 1) $0.\tt f$ is always kept as one of 256 ($8$-bit) quantized values. This is because $0$ has a special significance in CNNs such as zero padding. 2) By analyzing the histogram of parameters, we found most values are concentrated in a small range. This implies instead of using uniform quantization, we should put more codebooks for these concentrated region. The Lloyd-max quantization $\sf Quant(\cdot)$ is adopt to find the optimal codebooks $\{c_1, \ldots, c_{256}\}$ and bin-boundaries $\{l_1, \ldots, l_{257}\}$ \cite{scheunders1996genetic}. 3) When training deep neural networks, the parameters, activations and gradients have very different ranges. For example gradient's ranges slowly diminish during the training. Therefore, we draw the distributions and compute the codebook for each layer's weights, bias, activations and respective gradients separately. The training process is illustrated in Algorithm \ref{alg:train} and Figure \ref{fig:quant}.

\vspace{-0.2cm}
\subsubsection{Why not train on encrypted data}
\vspace{-0.2cm}
Note one important feature of this framework is that the DPN can be trained on unencrypted data and applied to encrypted data. In theory, it is possible 
to also train the DPN over encrypted data as its gradients are polynomials as well. 
However, in practice this is not scalable as it is very expensive to encrypt the entire training corpus and compute the gradient in encrypted domain. Quantization will also be an challenge if training on encrypted data. Besides, the learning algorithm does not have access to the secret key for decryption, we will never know what these trained weights are.

%\subsection{Decoding with Quantization} 

\vspace{-0.2cm}
\section{Experiments}
\label{sec:Exp}
\vspace{-0.2cm}

In this section we present the details of network architectures, practical considerations for training and decoding, and experimental results. All the models investigated in this work were trained using the computational network toolkit (CNTK) \cite{yu2014introduction}. The homomorphic encryption is implemented using the SEAL library \cite{sealcrypto}. 
We evaluate the effectiveness of the DPN proposed in Section \ref{sec:PolyNet} on
the Switchboard and Cortana voice assistant tasks.

In Switchboard task, we use 309hr training set and NIST 2000 Hub5 as test set.
The features used in this set up is 40-dimensional LFB with utterance-level CMN. The outputs of network are 9000 tied triphone states. We verified the polynomial approximation on two models, DNN and CNN.  The DNN is a 6-layer ReLU network with batch normalization and 2048 units on each layer. The CNN is a 17-layer VGG network \cite{saon2017english, zhang2016end}, including 3$\times{\tt conv}(3,3,96)$\footnote{${\tt conv}(3,3,96)$ means kernel size is $(3,3)$ and that layer has 96 kernels.}, max-pooling, 4$\times{\tt conv}(3,3,192)$, max-pool, 4$\times{\tt conv}(3,3,384)$, max-pool, followed by two dense layer with 4096 units and softmax layer. Both models use $[t-30, t+10]$ as the input context.
The LM used in this task is from \cite{yu2016deep}. The vocabulary size is 226k. The WERs of above DNN and CNN and the corresponding DPNs are shown in Table \ref{tbl:results_SWB}. All models are trained using CE criteria. We leave the sequence training \cite{zhang2015deep, vesely2013sequence} for these models as the future work.

% \usepackage{multirow}
\begin{table}[]
\begin{tabular}{|cc||cccc|}
\toprule
\multicolumn{2}{|c||}{WER in \%}               & 16-bit & 8-bit & 4-bit & 2-bit \\ \midrule
\multicolumn{1}{|l}{\multirow{2}{*}{DNN}}
    %\begin{tabular}[c]{@{}c@{}}DNN\\ $\downarrow$ \\ DPN\end{tabular}
    %}}
                            & quantization & 14.7\%  & 14.9\% & 16.6\% & 100.4\%      \\
\multicolumn{1}{|l}{}                           & + retrain  & --  & 14.7\% & 14.9\% & 30.3\%    \\  \midrule
%
%\multicolumn{1}{c|}{\multirow{2}{*}{DPN}}      &  DNN$\rightarrow$DPN        & 15.8\%  & --  & -- & --   \\
\multicolumn{2}{|l||}{DNN$\rightarrow$DPN (Alg. \ref{alg:train})}      & 15.8\% & 15.8\% & 16.1\% & 30.8\%   \\ \midrule \toprule
%
%%%%%%%%%%%%
\multicolumn{1}{|l}{\multirow{2}{*}{CNN}}       & quantization &  12.2\% & 12.7\% & 15.4\% &  --      \\
\multicolumn{1}{|l}{}                           & + retrain  & --   & 12.3\% & 12.7\%  & --   \\  \midrule
%\multicolumn{1}{c|}{\multirow{2}{*}{DPN}}       &  CNN$\rightarrow$DPN   & 13.0\%  & --  &  -- &  --  \\
\multicolumn{2}{|l||}{CNN$\rightarrow$DPN (Alg. \ref{alg:train})}      & 13.5\% & 13.6\% & 14.0\% & -- \\  \bottomrule
\end{tabular}
\caption{The results of DNN, CNN and DPN on switchboard. Note we could not make recurrent models work for homomorphic encryption.}
\label{tbl:results_SWB}
\end{table}

In Cortana voice assistant task, we used about 3400 hours US-English data in training and 6 hours data (5500 utterances) for testing. The features used in this setup is 87-dim LFB (including 29-dim static, $\Delta$ and $\Delta\Delta$) with utterance-level CMN. The networks used in this setup have the same structure as above, but with 9404 tied triphone states. Table \ref{tbl:results_Cortana} summarizes the WER and the average latency per utterance (including encryption, AM scoring, decryption and decoding) on this Cortana task. 

\begin{table}[h] \vspace{-0.4cm}
\begin{tabular}{ccc |c|c|c|}
\cline{4-6}
                          &        &       & \multicolumn{3}{c|}{avg. latency per utterance} \\ \cline{2-6} 
\multicolumn{1}{c|}{}     & 16-bit & 4-bit & encryption  & decryption  & overall \\ \toprule
\multicolumn{1}{|c|}{DNN} & 12.9\% & 13.4\% &    --     &    --     &    $177$ms      \\ \midrule
\multicolumn{1}{|c|}{DPN} & 14.8\% & 15.5\% &   $202$ms          &   $16$ms     &  $373$ms     \\ \bottomrule
%%%%%%%%%%%%
\end{tabular}
\caption{The performance of DNN and DPN on Cortana task. The overall latency of DPN includes encryption, AM scoring, decryption and decoding based on 4-bit model.}
\label{tbl:results_Cortana} \vspace{-0.4cm}
\end{table}

\vspace{-0.2cm}
\section{Conclusion and Future Work}
\label{sec:conclusion}
\vspace{-0.1cm}

The cloud-based SR service empowers users or third-parties to try state-of-art speech recognition easily in their own tasks. However, sending audios about personal or company internal information to the cloud, raises concerns about privacy. The main contributions of this work include: 1) a cloud-local joint decoding framework that enables privacy preserving speech recognition. 
It allows users to send their data in an encrypted form to ensure that their data remains confidential, at mean while the server can still do speech recognition on their behalf without knowing the content. 2) a deep polynomial network that can be trained efficiently on unencrypted data and make predictions over the encrypted speech in real time. 
 We illustrate the effectiveness of model and framework on the Switchboard and Cortana voice assistant tasks with acceptable performance degradation and latency increased comparing with the traditional cloud-based DNNs. 
 %
 %The privacy preserving machine learning is still an open problem and recently becomes very attractive .
%
Future works will include 1) making the decoder also work on encrypted domain so that we could move everything to the cloud. 2) investigating training on encrypted data so that multiple parties (e.g. Microsoft, Google and Amazon) can encrypt and combine their data together to train models without sacrificing users privacy. 

% Below is an example of how to insert images. Delete the ``\vspace'' line,
% uncomment the preceding line ``\centerline...'' and replace ``imageX.ps''
% with a suitable PostScript file name.
% -------------------------------------------------------------------------
% \begin{figure}[htb]

% \begin{minipage}[b]{1.0\linewidth}
%   \centering
%   \centerline{\includegraphics[width=8.5cm]{image1}}
% %  \vspace{2.0cm}
%   \centerline{(a) Result 1}\medskip
% \end{minipage}
% %
% \begin{minipage}[b]{.48\linewidth}
%   \centering
%   \centerline{\includegraphics[width=4.0cm]{image3}}
% %  \vspace{1.5cm}
%   \centerline{(b) Results 3}\medskip
% \end{minipage}
% \hfill
% \begin{minipage}[b]{0.48\linewidth}
%   \centering
%   \centerline{\includegraphics[width=4.0cm]{image4}}
% %  \vspace{1.5cm}
%   \centerline{(c) Result 4}\medskip
% \end{minipage}
% %
% \caption{Example of placing a figure with experimental results.}
% \label{fig:res}
% %
% \end{figure}

%\vfill
%\pagebreak

\vfill\pagebreak

\footnotesize{
\bibliographystyle{IEEEbib}
\bibliography{strings,refs}
}
\end{document}